\PassOptionsToPackage{pdfpagelabels=false}{hyperref}
\documentclass[a4paper,fleqn,usenatbib]{mnras}
\usepackage[T1]{fontenc}
\usepackage{ae,aecompl}
\usepackage{graphicx}	\usepackage{amsmath}	\usepackage{amssymb}	
\newcommand{\lcdm}{$\Lambda{\rm CDM}$}
\def\galaxia{{\sc galaxia}}
\newcommand{\kms}{\ifmmode \,\rm km\,s^{-1} \else $\,\rm km\,s^{-1} $ \fi }

\title[Clustering in the stellar halo of M31]{Architecture of the Andromeda: a quantitative analysis of clustering in the inner stellar halo}

\author[P. R. Kafle et al.]
{P. R. Kafle,$^{1}$ \thanks{E-mail: prajwal.kafle@uwa.edu.au}
S. Sharma,$^{2}$
A. S. G. Robotham,$^{1}$
G. F. Lewis,$^{2}$ and 
S. P. Driver$^{1}$
\\\\
{$^{1}$ ICRAR, The University of Western Australia,35 Stirling Highway, Crawley, WA 6009, Australia}\\
{$^{2}$ Sydney Institute for Astronomy, School of Physics A28, University of Sydney, NSW 2006, Australia}\\
}

\pubyear{2016}
\begin{document}
\label{firstpage}
\pagerange{\pageref{firstpage}--\pageref{lastpage}}
\maketitle
\begin{abstract}
We present a quantitative measurement of the amount of clustering present in the inner $\sim30$ kpc of the stellar halo of the Andromeda galaxy (M31). 
For this we analyse the angular positions and radial velocities of the carefully selected Planetary Nebulae (PNe) in the M31 stellar halo. 
We study the cumulative distribution of pair-wise distances in angular position and line of sight velocity space, and find that  the M31 stellar halo contains substantially 
more stars in the form of close pairs as compared to that of a featureless smooth halo. 
In comparison to a smoothed/scrambled distribution we estimate that the clustering excess in the M31 inner halo is roughly $40\%$ at maximum and on average $\sim 20\%$.
Importantly, comparing against the 11 stellar halo models of \cite{2005ApJ...635..931B}, which were simulated within the context of the \lcdm\ cosmological paradigm, 
we find that the amount of substructures in \emph{the M31 stellar halo closely resembles that of a typical \lcdm\ halo}.
\end{abstract}

\begin{keywords}
galaxies: individual (M31)- stars: individual (Planetary Nebulae)-methods: statistical
\end{keywords}

\section{Introduction}
Stellar haloes of galaxies are laboratories to test the process of galaxy formation and evolution. 
Under the current paradigm of hierarchical assembly in a Cold Dark Matter (CDM) cosmology it is believed that a galaxy, specifically its halo, 
grows continuously through the in-fall of smaller galaxies \citep{1978MNRAS.183..341W,1978ApJ...225..357S,1999MNRAS.307..495H}.
Observational evidence for the existence of a range of substructures in galaxy haloes is seen in tidal streams 
\citep[e.g.][etc]{1998Natur.396..549G,1998ApJ...504L..23S,2000MNRAS.314..324C}, stellar shells \citep[e.g.][etc]{1980Natur.285..643M,1990dig..book..270S} and fans \citep{1997ApJ...490..664W}.
In the local universe, the serendipitous discovery of tidal streams and substructures in the halo of the M31 and Milky Way galaxies 
\cite[e.g.][etc]{1993ARA&A..31..575M,1995MNRAS.277..781I} provide additional evidence of ongoing accretion.
The most prominent and spectacular among the contemporary findings are  the Field of Streams \citep{2006ApJ...642L.137B} in the Milky Way and the Giant stream \citep{2001Natur.412...49I} in M31. 

The amount of substructure in a given halo and the distribution of its properties, such as size, shape and location, are related to the accretion history of the halo.
In addition to accretion, the stellar halo is expected to be partly formed by in-situ stars.
Recent hydrodynamical simulations of galaxy formation  \citep{2006MNRAS.365..747A,2009ApJ...702.1058Z,2011MNRAS.416.2802F,2012MNRAS.420.2245M} suggest that in the inner regions, 
the stellar halo might be dominated by in-situ stars whose kinematic properties are distinct from accreted stars.
In the light of these findings, it is crucial to determine the relative contribution of the accreted component over that of an 
in situ component, for which a necessary first step would be to quantify the presence of observed substructures.
By studying clustering of stars in the stellar halo one can put constraints on the accretion history of that halo \citep{2008ApJ...689..936J,2011ApJ...728..106S}.

The two spiral galaxies in the Local Group, whose substructures can be studied in great detail through individual resolved stars, are the Milky Way and the neighbouring M31.
For the Milky Way there has previously been extensive effort to quantify the amount of clustering in the stellar halo 
e.g. \cite{1995MNRAS.275..429L,2006MNRAS.371L..11C,2008ApJ...680..295B,2009ApJ...698..567S,2011MNRAS.417.2206C,2011ApJ...728..106S,2011ApJ...738...79X,2016ApJ...816...80J} etc.
For M31, the relatively recent discovery of various substructures, 
such as G1 Clump \citep{2002AJ....124.1452F}, NE Clump \citep{2004ApJ...612L.117Z}, NE/W Shelves \citep{2002AJ....124.1452F} etc, and 
tidal streams \citep{2007ApJ...671.1591I,2009Natur.461...66M,2009MNRAS.396.1842R,2011ApJ...731..124C} also establish a qualitative assurance 
of clustering\footnote{for recent up-to-date menu of substructures in M31 refer to the \cite{2016ASSL..420..191F} in its stellar halo}.
However, to date there has only been a few attempts to quantify the level of substructure in the M31 stellar halo \citep{2009ApJ...701..776G,2012ApJ...760...76G,2014ApJ...780..128I}.
Here we present a new approach of estimating the amount of clustering in the M31 halo, as well as provide a comparative study with contemporary models.
 
Structures formed by accretion events exist in coherent phase space defined by their positional coordinates and velocity vectors.
Due to its large distance the proper motions of M31 halo stars are not known at a high enough fidelity to properly resolve all dimensions.
This leaves us with the angular position in sky (i.e. a galactic longitude and latitude; $l, b$) and a line of sight velocity ($v_{\rm los}$).
Although it is generally an arduous campaign $v_{\rm los}$ can be obtained from spectroscopic observations, and is accurate enough to resolve the halo substructures. 
Fortunately the bright emission lines of the Planetary nebulae (PNe), a population of dying stars, make them easily detectable and their line-of-sight velocities measurable with a minimal telescope time. 
Therefore, PNe have been proven to be excellent tracers for the dynamics of nearby galaxies leading to them being increasingly exploited to study the kinematic and dynamical 
properties of the M31 \citep[e.g.][etc]{1987ApJ...317...62N,2000MNRAS.316..929E,2001A&A...377..784N,2003ApJ...583..752E,2004ApJ...616..804H,2014AJ....147...16K} 
as well as some early-type galaxy haloes \citep[e.g.][etc]{2004ApJ...602..685P,2015A&A...579A.135L,2013MNRAS.436.1322C}.
To date the survey using the Planetary Nebulae Spectrograph by \cite{2006MNRAS.369..120M} and \cite{2006MNRAS.369...97H} provides the largest stellar sample in M31 with radial velocity information.
Unfortunately, many of these reside in the inner disk and bulge of the galaxy.
In general how representative PNe are of the whole halo population is still an open question.
Nonetheless, a few hundreds of PNe in \cite{2006MNRAS.369..120M} catalogue reside in the disk-halo interface or the inner halo of M31, which we use in this study.

In this paper, we provide a quantitative assessment of the amount of clustering present in the inner stellar halo of M31.
For this we utilise the angular coordinates and the radial velocity from archival data of Planetary Nebulae in M31 stellar halo.
The layout of this paper is as follows: in Section~\ref{sec:data} we describe the main properties of M31 and our sample of Planetary Nebulae (PNe).
Details of the method we undertake to quantify the substructures in M31 halo is presented in Section~\ref{sec:metric}.
In Section~\ref{sec:results} we present results of our findings and comparisons with the theoretical models, and further discuss them in Section~\ref{sec:discussion}.
In Section~\ref{sec:conclusion} we provide our concluding remarks.

\section{DATA}\label{sec:data}
\subsection{Central Properties: M31}
A list of the main properties of the M31 galaxy assumed in this work are as follows:
\begin{gather*}
\begin{cases}
\text{RA} & 00^{\text{hh}}42^{\text{mm}}44.33^{\text{ss}} \\
\text{Dec} & 41^{\circ}16'07.5'' \\
\text{Position angle}\ (\theta)^1       & 37.7^{\circ} \\
\text{Inclination angle}\ (i)^2         & 77.5^{\circ} \\
\text{Distance from MW}^3              & 780 \ \text{kpc}    \\
\text{Heliocentric radial velocity}^4  & -301 \ \kms \\
\end{cases}
\end{gather*}
$^1$ \cite{1958ApJ...128..465D}, 
$^2$ \cite{1992ASSL..176.....H},
$^3$ \cite{1998AJ....115.1916H,1998ApJ...503L.131S,2012ApJ...758...11C}, and
$^4$ \cite{1991rc3..book.....D,2008ApJ...678..187V}.
Observed heliocentric velocities are converted to Galactocentric ones by assuming a solar peculiar 
motion $(U_\odot,V_\odot,W_\odot)=(+11.1,+12.24,+7.25) \kms$ \citep{2010MNRAS.403.1829S} and a circular speed of $220 \kms$ at the 
position of the Sun\footnote{Note that we adopt the IAU recommended value but our main results are unchanged if we use higher 
value of $\sim240 \kms$ \citep[e.g.][etc]{2009ApJ...700..137R,2014ApJ...794...59K,2016arXiv160207702B}}. 
To convert a M31 rest-frame centred coordinate system to the heliocentric frame and vice-versa, 
we apply the formulary provided in the Appendix B of \cite{2007MNRAS.374.1125M}, 
mainly used for the coordinate conversion of the synthetic data discussed latter.

\subsection{Stellar Halo Tracers: Planetary Nebulae }\label{sec:datapne}
The intrinsic faintness and the vast spatial extension of substructures in 
M31's halo \citep[e.g.][etc]{2009Natur.461...66M,2011ApJ...732...76R,2012ApJ...760...76G,2014MNRAS.442.2165H,2014ApJ...780..128I} mean it is difficult to do a comprehensive survey of the stars therein. 
However, due to the development of wide-field multi-object spectrographs \citep{2002PASP..114.1234D} it is now easy to detect one of the brightest and ubiquitous halo tracer i.e. the Planetary Nebulae (PNe). 
As discussed previously, PNe census has surged lately and currently constitutes the largest stellar sample of M31 halo tracers with radial velocity information. 
Hence, we use PNe data in this work, which is obtained from the catalogues constructed by \cite{2006MNRAS.369..120M,2006MNRAS.369...97H}. 

The data were observed using the Planetary Nebula Spectrograph on the William Herschel Telescope in La Palma and were 
subsequently made publicly available online \footnote{\url{http://www.strw.leidenuniv.nl/pns/PNS_public_web/PN.S_data.html}}.
In total, the published catalogue provide 3300 emission-line objects of which only 2730 are identified as possible PNe.
Out of the likely-PNe sample roughly $6\%$ of objects are associated with M32, NGC 205, Andromeda IV or even external galaxies.
After removing above contaminants we are remained with an initial sub-sample of 2637 PNe belonging to M31.
Distances to these PNe are unfortunately unknown and typical uncertainty in their radial velocities is 20~$\kms$.
\begin{figure*}
   \centering
      \includegraphics[width=1.5\columnwidth]{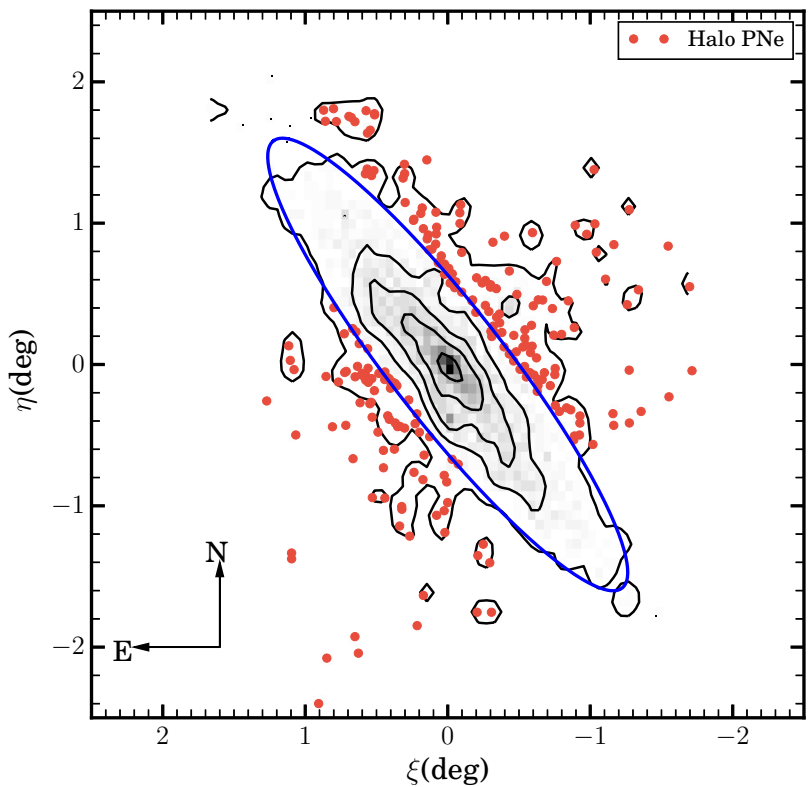}   \caption{M31 PNe data: Contours show an iso-density region of Planetary Nebulae in M31 (N=2637) whereas the red points denote its sub-sample (N=228) residing in the stellar halo.
   The halo PNe observation extends to a radius of $\sim30$~kpc.
   The ellipse shown with blue solid line represents a disk of inclination $77.5^\circ$ and radius $2^\circ$ (27.2 kpc), the approximate limit of the M31 visible disk.}
    \label{fig:data}
\end{figure*}

Our study only focuses on the tracers of the M31 stellar halo. 
Therefore, to select only objects dwelling in the halo we restrict the above sample of 2637 PNe 
to those with a projected distance of $|Y|\geqslant5$ kpc, the limit adopted from \cite{2008ApJ...674..886L}, 
which provides us the final sample of 228 halo PNe.
\cite{2008ApJ...674..886L} find that the  velocity dispersion of globular clusters system doubles beyond the projected distance of 5 kpc compared to the inner 1 kpc region, 
meaning pressure support dominates over the rotational support in the outer regions.
This provides an empirical proof that this cut delineates the halo-disk interface.
This can be seen clearly in Figure~\ref{fig:data}.
The contours show an iso-density region for our entire PNe sample associated with the M31 and similar to one in 
e.g. \cite{2002AJ....124.1452F,2009Natur.461...66M,2014ApJ...780..128I} who use red giant branch stars(RGBs).
The angles $\xi$ and $\eta$ are the conventional on sky angles, which are obtained from the gnomonic projection of the object's RA and Dec position.
The halo sub-sample obtained after the $|Y|\geqslant5$ kpc are shown with red dots.
The blue solid ellipse demarcates the approximate limit of the visible disk of inclination $77.5^\circ$ and radius $2^\circ$ (27.2 kpc).
Our halo PNe sub-sample all fall outside the ellipse means the $|Y|\geqslant5$ kpc cut is successful in selecting the halo candidates.
The observational extent of our final PNe data only extends to $\sim$ 30 kpc from the centre of the galaxy.
This represents only a $~1/4$ of the total stellar halo, which extends to at least 150~kpc \citep{2009Natur.461...66M,2012ApJ...760...76G}.
Thus, we are actually only probing the inner stellar halo of M31, or more strictly the disk-inner halo interface.

\section{Distance Metric and Interpretation}\label{sec:metric}
The main aim of our work is to quantify the level of clustering in M31 halo and subsequently compare this with predictions from theoretical models.  
A particularly useful way to quantity clustering in any data, consisting of points in some arbitrary space, is the two point correlation function.  
In astronomy this has been used extensively, mainly to characterise the spatial distribution of galaxies obtained from galaxy-redshift surveys 
\citep[e.g.][etc and references therein]{1980lssu.book.....P,1993ApJ...412...64L,2010gfe..book.....M}. 
Similarly, a correlation function based approach has also been used to study clustering in the context of the Milky Way galaxy \citep{2009ApJ...698..567S,2011MNRAS.417.2206C, 2011ApJ...738...79X}. 
We follow a similar approach, described below. 

Ideally, one would like a full phase space information to perform this kind of study.
However, in reality we only possess angular positions, e.g. galactic coordinates $l$ and
$b$, and radial velocity\footnote{basically, heliocentric coordinates with the contribution of the solar motion subtracted off the line-of-sight velocity} $v_{\text{los}}$ of tracers (our PNe).
The basic concept of this approach is to compute pair-wise distance of all tracers and then study the distribution of these distances. 
However, to compute the distances a proper choice of metric is required. 
Following \cite{2009ApJ...698..567S} and \cite{2011ApJ...738...79X}, we compute the separation between two PNe $i$ and $j$ using the following metric: 
\begin{equation}\label{eqn:metric}
s_{ij} = w_\theta \theta_{ij}^2 + w_{v_{\text{los}}} (v_{\text{los},i}-v_{\text{los},j})^2,  
\end{equation}
where angular separation 
\[\theta_{ij} = \cos^{-1} \{\cos b_i \cos b_j \cos(l_i-l_j) + \sin b_i \sin b_j\}. \]
The $w_\theta$ and $w_{v_{los}}$ are weights used to create a consistent metric for the two dimensions: position and velocity. 
They are derived from the reciprocal of an ensemble average of the separation in the respective dimensions given as
\[
w_\theta = \frac{1}{\langle \theta_{ij}^2 \rangle} \ \text{and}\ w_{v_{\text{los}}} = \frac{1}{\langle (v_{\text{los},i}-v_{\text{los},j})^2 \rangle}.
\]

A simple interpretation of Equation~\ref{eqn:metric} is that stars with small value for $s_{ij}$, or separation, are close-pairs and are likely to be a part of the same object or substructure. 
Furthermore, an excess of close-pairs, i.e., large number of stars with small $s_{ij}$, would mean more clustering.
On the contrary, a sub-structureless smooth halo, or a fully phase-mixed halo, should have fewer number of such close pairs. 

The performance of the above method has been well tested against various types of simulated data, e.g., 
on the sets of Monte Carlo samples drawn from simulations of disrupted satellites \citep{2009ApJ...698..567S}, on the data from 
semi-analytic model of galaxy formation \citep{2011MNRAS.417.2206C}, on the controlled simulation of Milky Way/M31 type galaxies \citep{2011ApJ...738...79X}.

The level of clustering can be further quantified using the information-theory based concept of Kullback-Leibler divergence \citep[KLD,][]{kullback1951information}.
The KLD measures the departure of the truth distribution $p$ from a candidate model $q$. 
For discrete probability distributions like the ones we get from Equation~\ref{eqn:metric}, the KLD of $q$ from $p$ on the set of points $i$ is defined as 
\begin{equation}\label{eqn:kld}
D_{\text{KL}}(p||q) = \sum_i p_i \log \left( \frac{p_i}{q_i} \right). 
\end{equation}
From the above formula we see that the KLD is an expectation value of the logarithmic difference between two probability distributions, computed with weights of $p_i$.
Also, $D_{\text{KL}}(p||q)$ is asymmetric in $p$ and $q$, and non-negative.
The greater the difference between the two distributions $p$ and $q$, the higher the value of $D_{\text{KL}}(p||q)$, and as $p \to q$, $D_{\text{KL}}(p||q) \to 0$. 
Here, both the distributions $p$ and $q$ are discrete and obtained from the histograms of a true close pair distribution and for the averaged distribution of the 200 random versions of the scrambled data.
Scrambled data mean the data with the intrinsic pairing between angles ($l,b$) and velocity ($v_\text{los}$) is decorrelated.
Scrambled data is assumed to represent featureless smooth data against which we can quantify the extent of clustering of the original sample.
A $D_{\text{KL}}(p||q) \geq 0 $ value means distribution of a true close pair is quantitatively different from the distribution of its randomised counterpart.

\section{Results}\label{sec:results}
Here we present and discuss the main results of the paper inferred from studying the distributions of pair-wise distance for a set of toy-data, 
the observed PNe data, as well as the synthetic data from simulations of the Milky Way type galaxies.

\subsection{Test data and a proof of concept}
\begin{figure}
   \centering
      \includegraphics[width=1.\columnwidth]{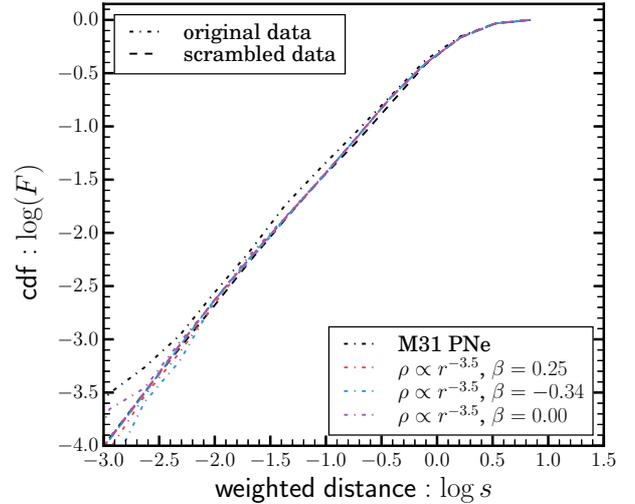}
      \caption[Close pair distribution]{Cumulative distributions of the pair-wise distances for PNe and synthetic data: 
   the black line shows the distribution for the observed M31 PNe data whereas the colourful lines show the distributions 
   for the featureless synthetic data with power-law density distribution ($\rho$) and a set of velocity anisotropies $(\beta)$.
   Dotted and dashed lines show the distributions before and after the angle and velocity pair were disassociated or scrambled.}
    \label{fig:cpd}
\end{figure} 

\begin{figure*}
   \centering
      \includegraphics[width=2\columnwidth]{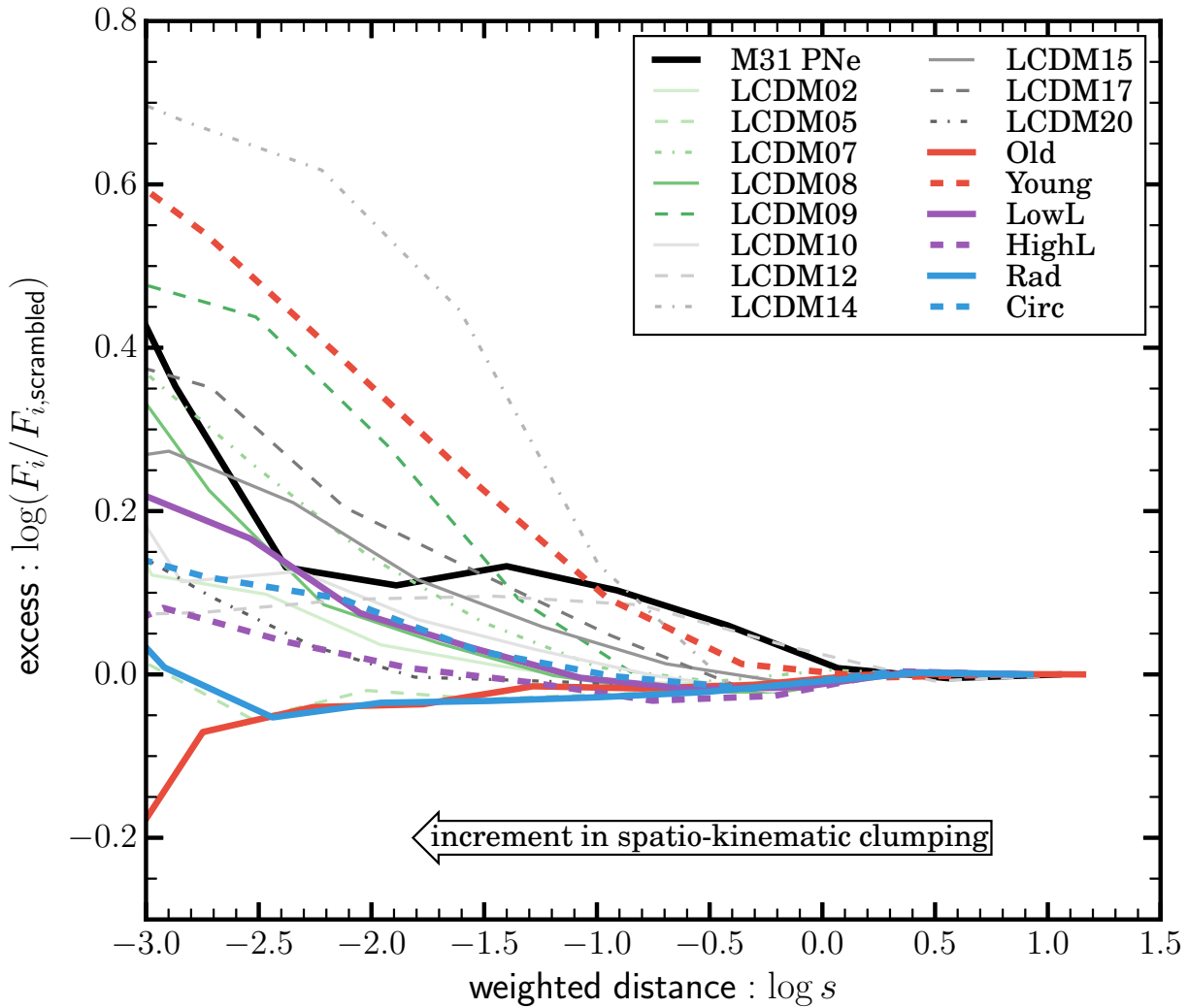}
      \caption[Clustering excess]{Clustering excess in the data compare to its scrambled case: a black line shows excess in clustering 
   for the observed PNe data whereas the remaining colours, as labelled in the figure, show excess in the 17 different \cite{2005ApJ...635..931B} 
   and \cite{2008ApJ...689..936J} simulated stellar haloes projected onto M31 observational space and convolved with observational errors.}  
    \label{fig:excess}
\end{figure*}
First, we demonstrate a proof of concept of our method for which we generate a set of test data.
The stellar volume density of M31 stellar halo is found to be a continuous single power-law i.e., $\rho \propto r^{-\alpha}$ with 
$\alpha\in[2.5,4]$ \citep[e.g.][etc]{2005ApJ...628L.105I,2007ApJ...671.1591I,2007ApJ...668..245G,2011ApJ...739...20C,2013ApJ...763..113D, 2014ApJ...780..128I}.
We generate particles following this power-law density distribution with $\alpha=3.5$.
To each of these particles we assign velocities that are sampled from either an isotropic ($\beta=0$) or a radial ($\beta=0.25$) or a tangential ($\beta=-0.34$) distribution.
$\beta$ is the Binney anisotropy parameter and describes an orbital structure of the system.
The values of $\beta$ can range from $-\infty$ (for purely circular trajectories) to 1 (for purely radial orbits).
From the above values of $\beta$ we obtain a set of 3 different test datasets, which we then project into an observational space `on sky'.
Henceforth, we treat them in the exact same way as we would the observed data.

In Figure~\ref{fig:cpd} we show the cumulative distributions of logarithm of the pair-wise distance obtained from Equation~\ref{eqn:metric} and the test data described above.
The red, blue and purple dotted lines show the distributions for the radial, tangential and isotropic  data respectively, whereas the dashed lines show the corresponding scrambled data. 
Precisely speaking, we scramble randomly the order of velocities but the order of angular positions is left unchanged. 
We then calculate the pairwise distance between two data points in this scrambled data.
To get more robust estimates we use 200 random versions of the scrambled data and only show their average distribution in the figure.

For all of the three datasets we measure $D_{\text{KL}}(p||q)\simeq0$, meaning the $\log F$ of the actual data and its scrambled counterpart are identical.
This result can also be attested visually from Figure~\ref{fig:cpd} by comparing the red/blue/purple dotted lines, 
which overlap the dashed lines (scrambled counterpart of the synthetic data) of the corresponding colour.
This is expected given that the synthetic data was featureless, or devoid of any substructure by nature. 
Similarly, we also find that changing the number density slopes ($\alpha$) to either a slightly higher value of 4 or a lower value of 2.5 still results $D_{\text{KL}}(p||q)\simeq0$.

Note that the black dashed line shows the level of clustering in the scrambled observed PNe data in M31 inner stellar halo. 
It overlaps with all the three featureless synthetic data.
This confirms that scrambling the real data satisfactorily produces a smooth distribution,
which we treat as a background and use to compare to any excess in clustering in the following section.
For a reference the black dotted line shows the level of clustering in observed PNe data in M31 inner stellar halo, which 
we further discuss and use in the following section.

\subsection{Observed data: PNe}
We now turn to quantifying the level of clustering in the M31 inner halo at $\lesssim 30$~kpc as traced by PNe.
Figure~\ref{fig:excess} shows the excess in clustering i.e. the logarithmic difference between the cumulative distributions of pair-wise distance, 
for the data and its scrambled counterpart, as a function of pair-wise distance in logarithmic scales.
In other words, it is the logarithmic difference between the black dotted and dashed lines shown in Fig.~\ref{fig:cpd}.
The result is shown with black solid line.
The smaller the $\log s$ value the closer the pairs are in spatio-kinematic space.
For these data we measure $D_{\text{KL}}(p||q) = 0.2 (\geqslant 0)$ meaning the cumulative distributions of the data and its scrambled counterpart are quantitatively different 
from each other i.e. there is an excess of clustering in the PNe data over a smooth distribution.  
Monte Carlo propagation of uncertainties in the radial velocities of the PNe is found to introduce less than a percent error in $\log F$ and $D_{\text{KL}}(p||q)$.
We estimate that the clustering excess of M31 inner halo compared to a smoothed distribution, i.e. with the zero excess, is roughly $40\%$ at maximum at the pair-wise 
distance $\log s = -3.0$ and on average $\sim 20\%$ over $\log s\in[-3.0,1.5]$.

\subsection{Comparison with simulated stellar haloes}\label{sec:simhalo}
To make theoretical predictions of substructures in the stellar halo, we use 17 stellar halo models of 
\cite{2005ApJ...635..931B} and \cite{2008ApJ...689..936J} and  generate stars from the models using 
\galaxia \footnote{The web-link to the \galaxia\ software is \url{http://galaxia.sourceforge.net/} and the 
original \cite{2005ApJ...635..931B} haloes were obtained from the web-link \url{http://user.astro.columbia.edu/~kvj/halos/}.
For the full documentation of the \galaxia\ refer to \url{http://galaxia.sourceforge.net/Galaxia3pub.html}} \citep{2011ApJ...730....3S}. 
Out of these simulated haloes, 11 (LCDM02 - LCDM20\footnote{for an easy reference the halo numberings are kept same as in the original catalogue}) 
have accretion histories derived from the \lcdm\ paradigm.  
The remaining six haloes have controlled/artificial accretion histories, with accretion events  that are predominantly (1) on radial orbits 
(ratio of the angular momentum of the orbit to the angular momentum of a circular orbit having same energy $\epsilon<0.2$), 
(2) on circular orbits ($\epsilon>0.7$), (3) old/ancient (time since accretion of greater than 11 Gyr), 
(4) young/recent (time since accretion of less than 8 Gyr), 
(5) high in luminosity (luminosity greater than $10^7 L_\odot$), and 
(6) low in luminosity (luminosity less than $10^7 L_\odot$) \citep[for further details see][]{2008ApJ...689..936J}.

To keep our simulated haloes consistent with the observed data, we set the \galaxia\ parameters such that it produces a catalogue of approximately 
the same sample size\footnote{determined by \emph{fSample} parameter in the \galaxia} as the observed PNe. 
Note, \galaxia\ samples an N-body system and if more than one star particle is spawned from an N-body particle, then this can introduce artificial smoothing on small scales. 
Generating fewer star particles avoids this problem.
We must choose a tracer population for our simulated stars and unfortunately, \galaxia\ does not generate PNe.
Hence, in instead we generated RGBs, for the discussion about this choice refer to Section~\ref{sec:discussion}.
To select RGBs from full halo population generated by \galaxia\ we use colour $B-V>0.8$ and absolute magnitude $M_V\leqslant4$.
To reduce the contamination of main sequence stars we have chosen a smaller value for $M_v$ compared to \cite{2000MNRAS.313..209H}. 
The stellar halo generated using \galaxia\ is then moved to the observational space of M31 following the transformation mentioned in Section~\ref{sec:data} 
i.e. the stars from the model are projected on the sky as the actual data.
Furthermore, we jitter their $v_\text{los}$ by a Gaussian random noise of $20 \kms$, which is similar to the magnitude of velocity uncertainties in the PNe data.
Uncertainties in the angular positions are assumed to be negligible.
Finally, we impose an additional restriction on the on-sky area of 4 square degrees to roughly match the observational extent of the PNe data.
However, we check that changing the on-sky area of the sample by $\pm1$ square degree has no discernible impact on our results.
We also mask the inner 2 square degrees region as done for the observed data (Section~\ref{sec:datapne}) earlier.
Also, the halo stars from the models are unbound from satellites of origin.

In Figure~\ref{fig:excess}, we show the clustering results of the above simulated haloes.
We plot clustering excess, which is the logarithmic difference between the cumulative distributions of pair-wise distance for the synthetic data and its scrambled counterpart.
Here again we only use $(l,b)$ and $v_\text{los}$ to calculate the pair-wise distance in exact same method as for the PNe.
As labelled in the figure different simulated haloes are shown with different colours. 
The solid and dashed lines of the same colour means haloes of opposite properties, e.g. old and young, radial and circular, and high luminosity and low luminosity. 
The grey and green lines with different line styles and shading show 11 \lcdm\ haloes (LCDM02 - LCDM20) whereas the black solid line show result with the PNe data.  

Reassuringly Figure~\ref{fig:excess} shows a number of expected trends for the simulated haloes. 
For example, the old halo shown with an orange solid line shows a less substructures than its counterpart young halo.
This is expected as substructures in an older halo get more time to phase mix. 
Similarly, the radial halo shows less substructure compared to the circular halo.
This is due to the fact that satellites on radial orbit pass close to the Galactic centre and are more likely to get disrupted due to strong tidal effects.  
The low and high luminosity haloes show similar clustering for $\log s > -1$, but at smaller $\log s$ low luminosity halo shows a systematic rise in clustering. 
The accretion time and the type of orbits are similar for both low and high-luminosity haloes, the main difference between them is the luminosity or mass of accretion events. 
The low luminosity halo has many accretion events but with small mass, and this generates small substructures that will only show up on smaller scales.
This explains the sharp rise in clustering at small $s$.  
On average the accretion history of a typical \lcdm\ halo is not as extreme as those of the six haloes discussed above.
As expected the clustering excess for the \lcdm\ haloes, except for the LCDM14, lies between the extreme accretion events. 
Overall, the above results confirms that our method successfully distinguishes the disparity in haloes.

Interestingly, the excess of substructures in the M31 PNe data is noticeably larger than the extreme smooth models such as the radial and old haloes.
In overall clustering in the M31 halo seems to be similar to that of a typical \lcdm\ haloes.
A similar result has also been observed in the Milky Way front by \cite{2008ApJ...680..295B}
and found the observations and simulations to be in rough agreement.
They also make use of the 11 simulated stellar haloes of \cite{2005ApJ...635..931B}.
However, they compared the radial dependence of the fractional root mean square deviation of 
the stellar density from a smooth triaxial model in the Galactic main-sequence turn-off stars instead.

\section{Discussion}\label{sec:discussion}
We note that the results presented here could be sensitive to some unquantified systematic biases.
Primarily, the PNe data we adopt from \cite{2006MNRAS.369..120M,2006MNRAS.369...97H} do not map the M31 stellar halo homogeneously.
The survey area mainly covers the disk of the galaxy (which we have masked out in this work) and a few fields beyond the optical radius of M31 (along the minor and major axes targeting in particular known substructures such as the Northern Spur and Southern Stream regions).
However, since no velocity limits have been imposed to obtain the PNe sample, we naively expect that the data we analysed have a reasonable completeness in a kinematic space.

As noted previously, \galaxia\ does not generate PNe data.
Hence, we compare our results for the observed PNe data against RGBs generated from the \galaxia\ model.
Our preference for RGBs is governed by two main motivations.
Firstly, RGBs are much fairer tracer of the stellar halo as they span a vast range in age and metallicity.
Therefore, they provide a good representation of the overall properties of the system in comparison to say blue horizontal branch stars, which represent very old stellar population.
Secondly, RGBs are typically the primary target in M31 studies \cite[refer to][and references therein]{2002AJ....124.1452F,2009ApJ...701..776G,2012ApJ...758...11C,2013ApJ...776...80M,2014ApJ...780..128I} mainly due to their intrinsic brightness and also, the ability to estimate their distances reliably.
Furthermore, we can see in Figure 29(30) of \cite{2006MNRAS.369..120M} that the PNe number counts along the major(minor) axis of M31 show excellent agreement with the r-band photometry, which is expected to be dominated by low-mass red giants.
As such, both RGBs and PNe trace similar intrinsic population and hence, their kinematic properties are expected to be similar.
However, we cannot discount the fact that considering different tracers could introduce systematic differences in the inferred properties of the accreted halo \citep{2011ApJ...728..106S}.
Hence, a slight fluctuation in the excess for the models shown in Figure~\ref{fig:excess} should be expected depending on the choice of tracer population.

Finally, we like to recap the studies by \cite{2009ApJ...701..776G,2012ApJ...760...76G} and \cite{2014ApJ...780..128I} which also quantify the amount of substructures in M31's halo by detecting kinematically cold components and over densities over a smooth halo distribution, respectively.
Most notably, \cite{2014ApJ...780..128I} find that $42\%$ of the most metal-poor populations in the M31 halo resides in the stream-like structures whereas metal-rich population are $86\%$ in streams.
Mainly due to differences in the approaches (i.e. statistical analyses) currently it is difficult to make a robust quantitative comparison of these works with our results.  
However, all these works qualitatively agree that a significant fraction of the stellar populations in the haloes of M31 are present in substructures.
Similarly, we can directly compare our results (black solid line in Figure~\ref{fig:excess}) for M31's halo to the equivalent studies of the MW halo in particular Figure 5. of \cite{2011ApJ...738...79X}.
Comparing this with our result for M31 we can say that the clustering excess of halo tracers in these two neighbouring galaxies are of a similar level.
For example, the excess of clustering for both the haloes at a pair-wise distance of $\log s=-2$ is roughly 0.15.
However, note we only probe the inner halo of M31 whereas \cite{2011ApJ...738...79X} probes the MW stellar halo out to a radius of 60~kpc.
Moreover, \cite{2011ApJ...738...79X} use a different halo tracer (blue horizontal branch stars) and also, use a three-dimensional spatial and velocity to measure the pair-wise distances.
Currently, due to different sample size, spatial coverage, type and extent of stellar halo tracers it is difficult to make a fair and robust comparison between the haloes of the MW and M31.

The ability of our method to distinguish the level of clustering in haloes with different accretion history, as shown in Section~\ref{sec:simhalo}, is encouraging.
In future, it will be useful to repeat the analysis with tracers that extend to much larger distances and far away from the disk, such as using the SPLASH \citep{2009ApJ...705.1275G} and PAndAS \citep{2009Natur.461...66M} surveys.
Such studies will allow statistically more robust quantification of clustering in the M31 halo and additionally, it will also open an avenue to produce a direct like-for-like comparison with existing measurements of clustering in the Milky Way stellar halo.

\section{Conclusions}\label{sec:conclusion}
We analyse archival data of Planetary Nebulae (PNe; \citealt{2006MNRAS.369..120M,2006MNRAS.369...97H}) 
to quantify the amount of kinematic clustering present in the inner $\lesssim 30$~kpc of the M31 stellar halo.
Additionally, we also compare our results with predictions from N-body simulations using the 17 synthetic stellar haloes of \cite{2005ApJ...635..931B} and \cite{2008ApJ...689..936J}.
 
In the three-dimensional space formed by angular coordinates and line of sight velocity, 
we find that the M31 stellar halo shows a significant amount of clustering, when measured in the form of excess of close pairs as compared to its scrambled counterpart.
We estimate that the clustering excess of M31 inner halo compare to a smoothed distribution
is roughly $40 \%$ at maximum (for the closest pairs) and $\sim 20 \%$ on average.

Furthermore, the study of stellar haloes of different accretion histories taken from \cite{2005ApJ...635..931B,2008ApJ...689..936J} models and generated using \galaxia\ 
software allows a quantitative calibration of our result with PNe.
We find that the excess of clustering in M31 is larger than the extreme smooth halo models such as the radial halo 
(one formed with accretion events that are predominantly on radial orbits) or the old halo (time since accretion of $>11$Gyr).
Importantly, we find that the clustering excess in \emph{the M31 inner stellar halo is similar to that of a typical \lcdm\ halo} in \cite{2005ApJ...635..931B} models.

\section*{Acknowledgements}
PRK acknowledges support from the Australian Research Council Discovery Project grant DP140100395 and Research Collaboration Awards 
(RCA) 12104401 and 12105203. We thank Magda Guglielmo, Anthony Conn, Pascal Elahi, Luke Davies, Manuel Metz and Marcel Pawlowski 
for their valuable time while preparing this paper. We are greatly thankful to the referee's constructive report that has helped improve the quality of the paper. 

This research made use of Astropy, a community-developed core Python package for Astronomy \citep{2013A&A...558A..33A}.
Also a sincere thanks to \cite{Hunter:2007}, \cite{PER-GRA:2007} and \cite{mckinney2012python} for their brilliant software, which were used extensively in this article. 
Figure~\ref{fig:data} makes use of \cite{2015ApJS..216...29B} utility package provided at \url{http://github.com/jobovy/galpy}.
We also benefited from J. T. VanderPlas \href{https://jakevdp.github.io/blog/2013/06/15/numba-vs-cython-take-2/}{blog post} on a pairwise distance function.

The Planetary Nebulae Spectrograph is a dedicated instrument for the study of the motions of planetary nebulae in galaxies. 
The project team involves astronomers from Groningen, Nottingham, Naples, MPE, ESO, Leiden and Mt. Stromlo. 
We acknowledge the team for making the PNe data available publicly in a timely manner.

\bibliographystyle{mnras}
\bibliography{paper}

\bsp	\label{lastpage}
\end{document}